\documentclass{sig-alternate-05-2015}
\usepackage[ruled, vlined, boxed]{algorithm2e}
\usepackage{algorithmic}
\usepackage{amsmath}
\usepackage{subfigure}
\usepackage{tabularx}
\usepackage{graphicx}
\usepackage{epstopdf}
\graphicspath{ {./fig/} }
\usepackage{epsfig}

\usepackage{etoolbox}
\usepackage{balance}

\begin{document}



%

\title{A new hierarchical clustering algorithm to identify non-overlapping like-minded communities}

%
%
%
%
%

\numberofauthors{5} 
%
\author{
%
%
\alignauthor
Talasila Sai Deepak\\
       \affaddr{Indian Institute of Technology, Guwahati}\\
       \email{t.deepak.iitg@gmail.com}
\alignauthor
Hindol Adhya\\
       \affaddr{Indian Institute of Technology, Guwahati}\\
       \email{hindol.adhya@gmail.com}
\alignauthor 
Shyamal Kejriwal\\
       \affaddr{Indian Institute of Technology, Guwahati}\\
       \email{shyamalkejriwal@gmail.com}
\and  
\alignauthor Bhanuteja Gullapalli\\
       \affaddr{Indian Institute of Technology, Guwahati}\\
       \email{g.bhanu@iitg.ernet.in}
\alignauthor Saswata Shannigrahi\\
       \affaddr{Indian Institute of Technology, Guwahati}\\
       \email{saswata.sh@iitg.ernet.in}
}
\date{30 Jan 2016}

\maketitle
\begin{abstract}
A network has a non-overlapping community structure if the nodes of the network can be partitioned into disjoint sets such that each node in a set is densely connected to other nodes inside the set and sparsely connected to the nodes outside it. There are many metrics to validate the efficacy of such a structure, such as clustering coefficient, betweenness, centrality, modularity and like-mindedness. Many methods have been proposed to optimize some of these metrics, but none of these works well on the recently introduced metric like-mindedness. To solve this problem, we propose a behavioral property based algorithm to identify communities that optimize the like-mindedness metric and compare its performance on this metric with other behavioral data based methodologies as well as community detection methods that rely only on structural data. We execute these algorithms on real-life datasets of Filmtipset and Twitter and show that our algorithm performs better than the existing algorithms with respect to the like-mindedness metric.
\end{abstract}



    \section{Introduction} \label{sec:introduction}

    Networks present a natural way of representing complex systems having many components and inter-component relationships. A network may represent a social, biological or a technological system. Networks in real life display certain properties such as the power law of degree distribution and small world phenomenon. Another such property is the \emph{community structure}, i.e., the presence of either overlapping or non-overlapping groups of nodes having dense connections internally. There are many algorithms in the literature for finding non-overlapping communities, e.g., the hierarchical clustering algorithms \cite{avelink, clink, graphlink, slink}, Girvan-Newman algorithm \cite{newman2004finding} and Louvain method \cite{louvain} for modularity maximization. Most of these algorithms exploit the structural properties of a network, such as the existence of an edge between two particular vertices. However, there can be many \emph{behavioral properties} associated with each node in a network, such as purchasing habits in transactional data, and in the context of our experiments, movie ratings or celebrity following by the users in an online social network. Modani et al. \cite{modani2012like} recently presented a method to identify an overlapping community structure having high like-mindedness. Even though this method works well in practice, it does not give a theoretical upper bound on its worst-case running time. Moreover, the efficacy of this method has not been evaluated with respect to structural metrics like modularity.  In this paper, we present a new algorithm \emph{Like-mindedness Maximization} which achieves a high value of the goodness metric \emph{like-mindedness} in the non-overlapping community structure it finds. We compare this algorithm with other existing behavioral and structural data based non-overlapping community finding algorithms and show that it outperforms all of these on this metric on Filmtipset and Twitter data sets. Moreover, we modify an existing structural based algorithm \emph{Louvain method} and supplement it with behavioral data in its input to compare its performance with our algorithm on both metrics \emph{like-mindedness} and \emph{modularity}. We observe that our algorithm achieves higher like-mindedness than what is obtained after this modification of Louvain method as well. We also observe that the modularity achieved by our algorithm is not significantly lower than what is obtained by the algorithms that only use structural properties of a graph with the objective of modularity maximization. The worst-case running time of our algorithm is $O({|V|}^2 \log |V|)$, which is slower than the best-known hierarchical clustering algorithms only by a factor of $O(\log |V|)$.

{\bf Definitions and Notations:} At this point, let us introduce the definitions and notations that we use in this paper. A network is denoted by an undirected and unweighted (i.e., each edge having weight $1$) graph $G = (V, E)$ with vertex set $V= \{1, 2, \ldots, |V|\}$. Since the graph represents a social network, we assume that the graph is sparse, i.e., $|E| = O(|V|)$. Each vertex $v \in V$ is associated with a behavioral vector $X_v$ of dimension $d$. The behavioral vector and its dimension depend on the particular dataset being used. For example, the ratings given by a user on a movie rating website (with some default rating being given to those movies he has not rated) can be his behavioral vector, the dimension of which is the number of movies available for rating. A similarity metric $sim(u, v)$ is a distance measure between the vectors $X_u$ and $X_v$ representing the behavior of the vertices $u, v \in V$, respectively. Examples of such distance measures are \emph{cosine similarity}, \emph{Manhattan distance}, \emph{Euclidean distance} and \emph{squared Euclidean distance}. Throughout this paper, we use cosine similarity $\frac{X_u \cdot X_v}{\|X_u\| \|X_v\|}$ as the distance measure.

Let $C = \{ C_{1}, C_{2}, \ldots, C_k \}$ be a partition of $V$, each representing a set of vertices (community) in a community structure having $k$ non-overlapping communities. 

{\bf Modularity} \cite{newman2004finding} is a well-known measure to capture the notion that a good community structure should have denser connections inside the communities, compared to the connections in-between communities. The modularity $Q(C)$ of the set of communities $C$ is defined as $Q(C) = \sum_{i=1}^k \left(a_{i} - b_{i}^{2} \right)$, where $a_{i}$ is the fraction of $|E|$ edges with both its vertices in the same community $C_i$ and $b_{i}$ is the fraction of $|E|$ edges with at least one vertex in community $C_i$. 

{\bf Like-mindedness} \cite{modani2012like} is a recent measure to capture the notion that members of a good community should share similar type of interests. The {\it like-mindedness} $L(C)$ of the set of communities $C$ is defined as the average of all intra-community vertex pair similarities. In other words, 
    \begin{equation}
    \nonumber 
    L(C) = \frac{1}{\sum\limits_{u, v \in V : u \leq v} \delta(u, v)} \sum\limits_{u, v \in V : u \leq v} sim(u, v) \cdot \delta(u, v)
    \end{equation}, where the boolean function $\delta(u, v)$ is $1$ if and only if there exists a community $C_i \in C$ such that $u, v \in C_i$.

\section{Structure based Clustering Algorithms}
In this section, let us briefly review the known structural property based clustering algorithms to detect non-overlapping communities. These methods only take the structural data into account, while deciding the optimal community structure for a given network. We evaluate these algorithms against our proposed algorithm with respect to both the metrics like-mindedness and modularity.

\begin{algorithm}
Calculate edge betweenness for all edges in the network. \\
\While{no edges remain}{
Remove the edge with highest edge betweenness. \\
Recalculate edge betweenness for all edges affected by the removal.}
\caption{Girvan Newman Algorithm}
\end{algorithm}

\begin{algorithm}
        Initialization: Assign each node to its own community.\\
        \While{modularity is improved}{
            \For{each $i$ $\in$ $V$}{
                \For{each $j$ $\in$ $neighbors(i)$}{
                    Calculate modularity gain by shifting $i$ to $j$'s community.
                }
            }
            Place $i$ in community for which gain is maximum.
        }
        \caption{Louvain Method}
    \end{algorithm}

\label{structuralalgorithms}
\subsection{Girvan-Newman Algorithm}
Girvan and Newman~\cite{girvan2002community} proposed a top-down algorithm in which they focused on edges that are the least central, i.e., the edges belonging to $E$ that are most ``between" communities. The \emph{edge betweenness} of an edge $e \in E$ is defined as the sum of the weighted number (defined below) of shortest paths passing through $e$. If there are $k$ distinct shortest paths between $u$ and $v$ for a pair of vertices  $u, v \in V$, each of the shortest paths connecting them is counted as $1/k$. By progressively removing the edge $e \in E$ having the highest edge-betweenness among all edges at the time of its removal, one is left with the communities as connected components of the graph. The running time of this method is  $O(|V| {|E|}^2)$, i.e., $O({|V|}^3)$ on a sparse graph.

\subsection{Modularity Maximization}
\label{sec:modularity_maximization}

Given a graph, finding a partition that maximizes the modularity is known to be an NP-hard problem \cite{brandes}. As a result, all practical algorithms for modularity maximization employ some approximate optimization methods such as greedy algorithm, simulated annealing or spectral optimization. One such example is the Louvain method \cite{louvain}, which is the most popular and fastest method for modularity maximization. It is a greedy algorithm in which we start with a community structure such that each node is its own community. In each step, we identify if shifting a vertex $i$ to any of its neighbor's community leads to an increase in modularity. If there is such an increase, we place $i$ in the community for which the increase is maximum. The steps are executed till there is no further improvement possible by shifting any vertex to any of its neighbor's communities. The running time of this method is estimated to be $O(|V| \log |V|)$ on sparse graphs.

The method of using modularity maximization to identify a community structure has a few drawbacks. First, it suffers from resolution limit~\cite{fortunato2007resolution}, i.e., as the network size increases, the ability to identify smaller communities by this method decreases. Secondly, there may be a huge number of partitions, each very different from the others, having high but similar modularity values, all close to the absolute maximum \cite{good2010performance}.
 
\subsection{Modified Louvain method}

As mentioned before, both Girvan-Newman algorithm and Louvain Method only use structural properties of a graph while detecting communities. As a 
result, these algorithms may not perform well with respect to the like-mindedness metric. Motivated by this, we present a slightly modified version of Louvain method in which behavioral data is used in addition to structural data to identify communities.

    \label{algo:mod_louv}
\begin{algorithm}
        Initialization: Assign each node to its own community. \\
        \While{modularity w.r.t. $G'$ is improved}{
            \For{each $i \in V'$}{
                \For{each $j$ $\in$ $neighbors(i)$ in $G'$}{
                    Calculate modularity gain w.r.t. $G'$ by shifting $i$ to $j$'s community.
                }
            }
            Place $i$ in community for which gain is maximum. \\
            Identify pairs of vertices $\notin E'$ having similarity greater than or equal to the like-mindedness of the current set of communities; add such pairs to $E'$.
        }
        \caption{Modified Louvain method}
\end{algorithm}
We start with an identical copy $G'=(V, E')$ of the graph $G = (V, E)$. In each iteration of the Louvain method, the idea is to shift a node from its own community to another community to gain modularity. After the shift of a node, the community structure changes along with its like-mindedness. At this point, there might be pairs of nodes $\notin E'$ having equal or higher similarity than the current like-mindedness of the community structure. In the modified Louvain Method that we propose, we add each such pair to $E'$ as additional edges. The intuition behind such addition of pairs to $E'$ is to increase the like-mindedness of the community structure in the future iterations since such additional edges are likely to help the placement of a node to a community having similar nodes. However, it can be noted that the inclusion of such additional edges might lead to a drop in modularity of the set of communities with respect to the given network $G$.

\section{Our Algorithm: Like-mindedness Maximization}

\begin{table}
\caption{Summary of linkage criteria commonly used by hierarchical clustering algorithms to decide whether to merge two communities $X$ and $Y$}
\label{tab:clustering_linkages}
\centering
\scalebox{0.9}{
\begin{tabular}{|c|c|}
\hline
\textbf{Linkage criteria} & \textbf{Formula}\\
\hline
Single linkage & $min\{sim(u, v) : u \in X, v \in Y \}$\\
\hline
Average linkage & $\frac{1}{|X||Y|} \sum_{u \in X} \sum_{v \in Y} sim(u, v)$\\
\hline
Complete linkage & $max\{sim(u, v) : u \in X, v \in Y\}$\\
\hline
\end{tabular}}
\end{table}

In this section, we present our algorithm to optimize like-mindedness. Our algorithm is a bottom-up hierarchical clustering approach 
in which one starts with having each vertex belonging to $V$ as its own community. In each subsequent step, pairs of communities are merged till there is only one cluster left. The pair of communities that gives the minimum value of a pre-defined linkage criterion (defined below) is selected for being merged in each step. In this agglomerative approach, we get a hierarchy of communities, often visualized as a dendrogram. 

There are a variety of hierarchical clustering algorithms \cite{slink}, \cite{clink}, \cite{avelink}, \cite{graphlink} depending on two parameters: a {\it similarity metric} between each pair of vertices and a {\it linkage criterion}. As mentioned before, we use cosine similarity as the similarity metric in this paper.
The linkage-criterion defines the similarity between a pair of communities $A$ and $B$ as a function of the pair-wise similarities of the members of the communities. In Table \ref{tab:clustering_linkages}, the three linkage criteria commonly used by hierarchical clustering algorithms are summarized. Note that each of these requires $O({|V|}^2)$ time to produce the complete hierarchy of communities. In the following paragraph, we give a new linkage criterion to determine the pair of communities to be merged at any step.

\label{newalgorithms}
    \vspace{-0.5em}
    \begin{algorithm}
         Initialization: Assign each node to its own community. \\
        \While{there is more than one community}{
            Merge the pair of communities $\{C_i, C_j\}$ having the highest score $S(C_i, C_j)$.
        }
        \caption{Like-mindedness Maximization (LMM)}
        \label{algo:lmm}
\end{algorithm}
We observe from the dendograms (not shown here due to space constraints) of hierarchical clustering algorithms that an algorithm produces higher like-mindedness if small clusters are merged in the early iterations in order to avoid the creation of large heterogeneous communities. Motivated by this fact, we design an algorithm in which we discourage the merging of two large communities at every iteration. At every iteration, we identify the pair of communities 
$\{C_i, C_j\}$ with the highest score $S(C_i, C_j)  = \frac{1}{max\{|C_i|, |C_j|\}} + \frac{1}{|C_i||C_j|} \sum_{u \in C_i, v \in C_j} sim(u, v)$. The left term is used to discourage the merging of two large communities, and the right term accounts for the average like-mindedness of the inter-community pairs from $\{C_i, C_j\}$. A high value of the right term ensures that the like-mindedness of the set of communities after merging is high,
since $\sum_{u \in C_i, v \in C_j} sim(u, v)$ is the sum of the similarities of $|C_i||C_j|$ pairs of inter-community vertices belonging to $C_i$ and $C_j$.


To implement the algorithm, we index each community by its smallest numbered node and maintain a $|V| \times |V|$ matrix $M$ that stores the pointers to the scores $S(C_i, C_j)$ of the respective pairs of communities $\{C_i, C_j\}$. To begin with, a max-heap $H$ with $O({|V|}^2)$ nodes is created in $O({|V|}^2)$ time to store the scores. When the pair of communities $\{C_i, C_j\}$ with the highest score has been identified from the heap and merged into $C_k$, we need to update the heap entries corresponding to $\{C_i, C_k\}$ for all $i \neq k$ and $\{C_j, C_k\}$ for all $j \neq k$. The update operation in a node requires the respective score to be recalculated, which takes a total of $O({|V|}^2)$ time over all iterations due to the fact that the similarity of a pair of vertices is used for the score calculation only when they belong to two different communities being merged. Since pairs of communities are merged in $O(|V|)$ iterations, each of which requires $O(|V|)$ update operations in a max-heap requiring $O(\log |V|)$ time per operation, the running time of the algorithm is $O({|V|}^2 \log |V| + {|V|}^2)$ $= O({|V|}^2 \log |V|)$.

\section{Dataset \& Experimental Setup} \label{sec:dataset_and_setup}

    \subsection{Filmtipset}

    Filmtipset is Sweden's largest movie rating website, in which a user has the option to rate a movie on a scale of $1$ to $5$. Apart from this, there is a social network element of the website where a user can {\it follow} another user in the network. We have $86,725$ such following relationships between the users. We designate two users $u, v \in V$ as friends if $u$ is following $v$, and vice-versa.

To use the dataset in our experiments, we apply a couple of filters. The first filter is applied on the number of times a movie has been rated. Some movies are {\it popular} since they are rated by many. We focus on the unpopular movies as an indication of the behavior of a user in this network, since a user is likely to know about an unpopular movie from the activities of his friends on filmtipset. We also observe that removing the most popular movies from being considered results in a higher ratio of the average similarity (w.r.t. rating vectors defined below) of the pairs of friends to the average similarity of the non-friend pairs. We denote this ratio for a graph $G$ as its {\it homophily ratio} $H(G)$. 

\begin{equation}
    \nonumber 
    H(G) = \frac{\frac{\sum\limits_{u, v \in V : (u, v) \in E} sim(u, v)}{|E|}} 
    {\frac{\sum\limits_{u, v \in V :  (u, v) \notin E} sim(u, v)}{{|V| \choose 2} - |E|}} 
    \end{equation}

It can be noted that aiming for too high a homophily ratio reduces the number of movies a lot. Since the number of movies left is used for filtering out inactive users (see below), a high homophily ratio implies a reduction in the number of active users as well. Therefore, we remove movies that are rated at least $50$ times in filter $1$ since it ensures a large number of movies left after filtering. To see the results of our experiments on networks having high homophily ratio, we remove movies that are rated at least $5$ times in filter $2$.

For each of the movie filters $1$ and $2$, we define a user filtering criterion as follows. For movie filter $1$ (filter $2$), we say a user to be {\it active} if he rates at least $5$ movies among the movies left after removing all movies rated more than $50$ ($5$, respectively) times. A user is called {\it social} if he has at least $5$ friends in the network. For each of movie filter $1$ and $2$, we create an induced subgraph such that each user in this subgraph is active and social. Here, it is worth reminding that movie filtering is always done before user filtering.  In Table \ref{filter_filmtipset}, we summarize the properties of these datasets, which we would denote by Filmtipset Filtered Networks 1 and 2 from now onwards. We also note that the Unfiltered Network as well as the Filtered Networks follow a {\it power law degree distribution}.

For either of Filmtipset Filtered Network 1 and 2, we consider the movies in an order and each user $u$ is assigned a {\it rating vector} $R_u$, 
each entry of which is either the rating given by him to that particular movie or $0$ if he has not rated it. We also create another {\it (un)interested vector} $S_u$ for each user $u$, each entry of which is $1$ or $0$ depending on whether a user has rated that movie or not, respectively. In order to implement the behavioral property based community finding algorithms, either $R_u$ or $S_u$ is used as the behavioral vector $X_u$ of $u \in V$.

\begin{table*}
\centering
        \caption{Comparison of running time on sparse graphs (graphs in which $|E|=O(|V|)$)} 
        \label{algo:time}
	\renewcommand{\arraystretch}{1}
	{
            \begin{tabular}{|c|c|c|c|c|c|c|c|}
                \hline
                  LMM & S \cite{slink}& A \cite{alink}& C \cite{clink} & GN & L (estimated) & ML (estimated)\\
                \hline
                  $O({|V|}^2 \log |V|)$ & $O({|V|}^2)$  & $O({|V|}^2)$  & $O({|V|}^2)$  & $O({|V|}^3)$ & $O(|V|\log |V|)$ & $O({|V|}^2 \log |V|)$\\
                \hline
            \end{tabular}
        }
	\end{table*}

    \begin{table}
    \caption{Properties of Filmtipset dataset}
    \label{filter_filmtipset}
    \centering
\renewcommand{\arraystretch}{1}
    \scalebox{0.75}{
    \begin{tabular}{|c|c|c|c|}
    \hline
    \textbf{Parameter} & \textbf{Unfiltered} &  \textbf{Filtered } & \textbf{Filtered}\\
    \textbf{} & \textbf{Network} &  \textbf{Network $1$} & \textbf{Network $2$}\\
    \hline
    Number of nodes & 91530 & 4305 & 983\\
    \hline
    Number of isolated nodes & 61211 & 168 & 152\\
        \hline
    Edge count (friendships) & 56387 & 10940 & 1807\\
    \hline
    Avg. clustering coefficient & 0.467 & 0.434 & 0.338\\
    \hline
    Avg. degree & 1.232 & 5.082 & 3.676\\
    \hline
    Diameter & 20 & 18 & 16\\
    \hline
    Avg. path length & 7.508 & 5.796 & 4.817\\
    \hline
    Size of giant component & 29.54\% & 90.77\% & 72.94\% \\
    \hline
    Homophily ratio & Not calculated & 26.44 & 62.07 \\
    \hline
    \end{tabular}}

    \end{table}

                \begin{table}
                \centering
                \caption{Properties of Twitter dataset}
                \label{filter_twitter}
\renewcommand{\arraystretch}{1}
	{
	\scalebox{0.7}{
                \begin{tabular}{|c|c|c|}
                \hline
                \textbf{Parameter} & \textbf{Unfiltered Network} & \textbf{Filtered Network 1}\\ \hline
                Number of Nodes & 40096646 & 5013 \\ \hline
                Zero degree Nodes & 17522652 & 1\\ \hline
                Edge count & 232157703 & 1636971 \\ \hline
                Avg. degree & 11.5799 & 653.09 \\ \hline
                Avg. Clustering coefficient & Not Calculated & 0.2063612 \\ \hline
                Diameter & 18 \cite{Kwak10www} & 4 \\ \hline
                Avg. path length & 4.12 \cite{Kwak10www} & 1.874073\\ \hline
                Size of giant component & Not Calculated & 5012 \\ \hline
                Homophily ratio & Not Calculated &  1.18 \\ \hline
                \end{tabular}}
        }

        \end{table}
        
        \begin{table}
\centering 
        \caption{Highest Modularity achieved by an algorithm}
        \label{mod_values}
	\renewcommand{\arraystretch}{1}
	\centering
	{
	\scalebox{0.75}{
		\begin{tabular}{|c|c|c|c|c|c|c|}
        \hline
        Fimltipset & LMMS & LMMR & GN & SS& AS & CS \\
        \hline
        Filtered Network 1 & 0.107 & 0.118 & 0.738 & 0.101 & 0.176 & 0.111 \\
        \hline
        Filtered Network 2 & 0.111 & 0.154 & 0.627 & 0.143 & 0.196 & 0.115 \\
        \hline 
        \hline
        Filmtipset & SR & AR & CR & LS &  MLS & MLR\\
        \hline
        Filtered Network 1  & 0.082 & 0.162 & 0.099 & 0.762 & 0.582 & 0.593\\
        \hline
        Filtered Network 2  & 0.141 & 0.204 & 0.1 & 0.676 & 0.614& 0.611\\
        \hline      
        \hline
        Twitter & A & C & S & LMM& L & ML\\
        \hline
        Filtered Network & 0.095 & 0.107& 0.004 & 0.078& 0.182 & 0.147\\
        \hline
		\end{tabular}}
        }
	\end{table}

\begin{table}
\centering
        \caption{Symbols used in Figure \ref{likemindedness}, and Tables \ref{mod_values} and \ref{algo:time}}
	\label{table:symbols}
	\renewcommand{\arraystretch}{1}
	\centering
	\scalebox{0.75}{
	\begin{tabular}{|c|c|}
	\hline
	\textbf{Symbol} & \textbf{Full form}\\
	\hline
	LMM/LMMS & LMM Algorithm using (un)interested vector\\
	\hline
	LMMR & LMM Algorithm using rating vector\\
	\hline
	L & Louvain method\\
        \hline
	ML/MLS & Modified Louvain method using (un)interested vector\\
        \hline
        MLR & Modified Louvain method using rating vector\\
	\hline
	GN & Girvan-Newman algorithm\\
	\hline
	S/SS & Single-linkage Clustering using (un)interested vector\\
	\hline
	A/AS & Average-linkage Clustering using (un)interested vector\\
	\hline
	C/CS & Complete-linkage Clustering using (un)interested vector\\
	\hline
	SR & Single-linkage Clustering using rating vector\\
	\hline
	AR & Average-linkage Clustering using rating vector\\
	\hline
	CR & Complete-linkage Clustering using rating vector\\
        \hline
        \end{tabular}
        }
	\end{table}

\subsection{Twitter} 

Twitter is one of world's largest micro-blogging website, in which a user has an option to follow another user. We use the publicly available dataset \cite{Kwak10www} which has the data of about $40$ million users. All the people having more than $10,000$ followers are designated as {\it celebrities}. 

For our experiments, we create a friendship graph called Twitter Filtered Network of the non-celebrity users who have at least $5000$ non-celebrity friends. In Table \ref{filter_twitter}, we summarize the dataset. We also note that the Twitter Filtered Network follows a \emph{power law degree distribution}. 

As before, two users $u$ and $v$ are said to be friends if $u$ is following $v$, and vice-versa. We create a $0/1$ vector $F_u$ (the $i$-th entry of which corresponds to the $i$-th celebrity) for each user $u \in V$ and use it as his behavioral vector $X_v$. The entry in $i$-th position of $F_u$ for $u$ is $1$ or $0$ depending on whether $u$ is following the $i$-th celebrity or not, respectively.

  \vspace{-1em}  
    \section{Results and Conclusions} \label{results}

\begin{figure*}[t]
{
{{\resizebox{2.25in}{1.95in}{\includegraphics{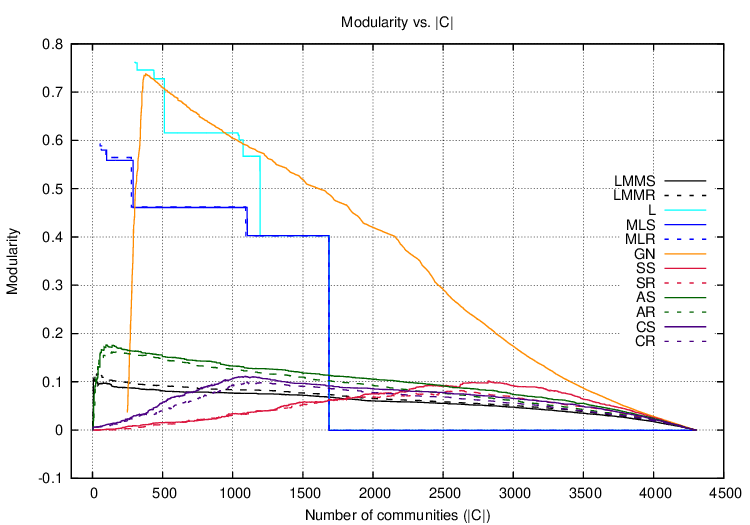}}}}
{{\resizebox{2.25in}{1.95in}{\includegraphics{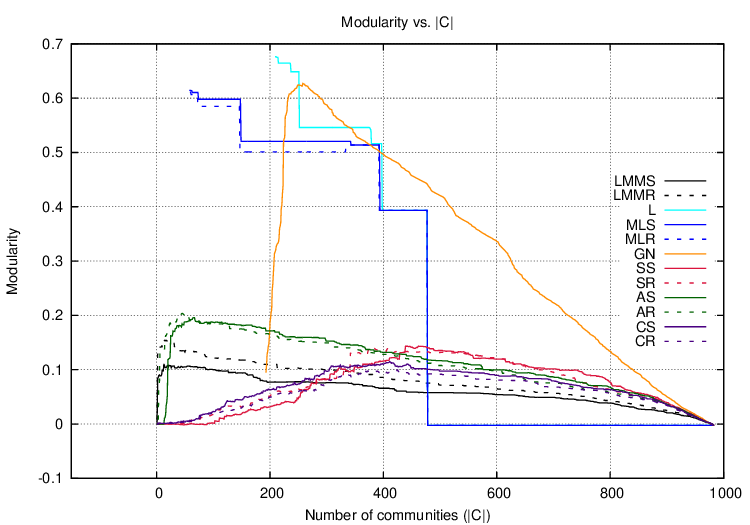}}}}
{{\resizebox{2.25in}{1.95in}{\includegraphics{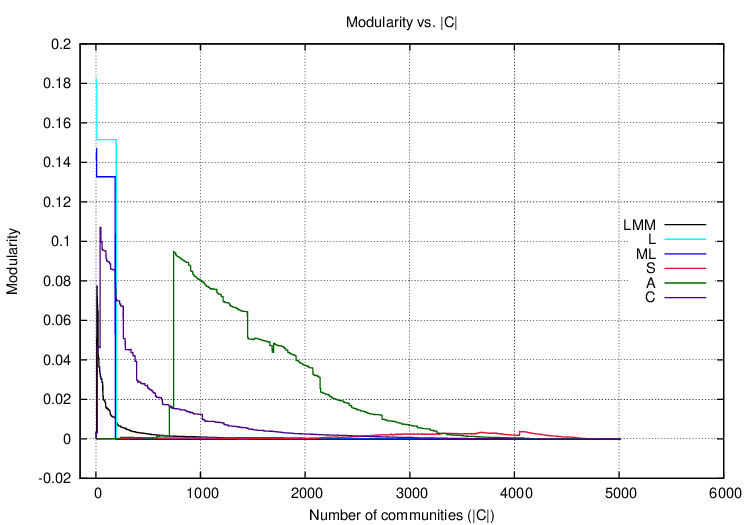}}}}
{{\resizebox{2.3in}{1.95in}{\includegraphics{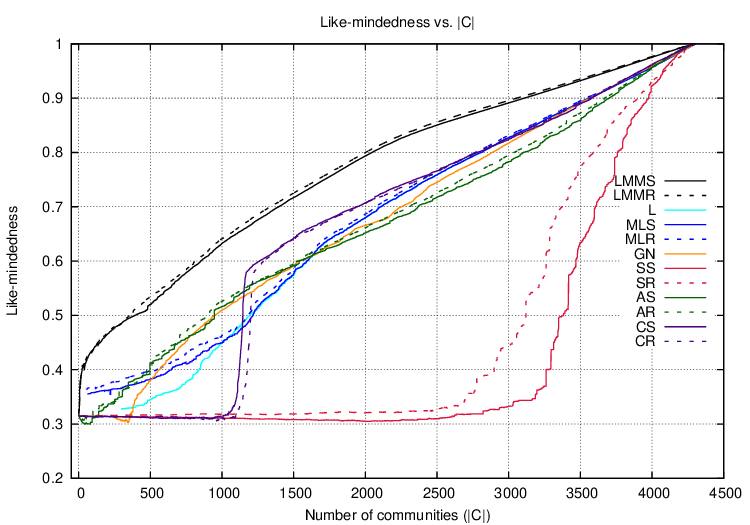}}}}
{{\resizebox{2.3in}{1.95in}{\includegraphics{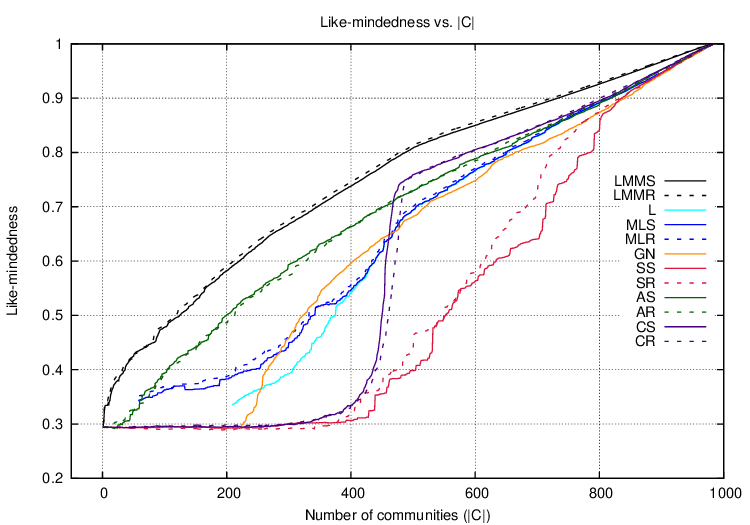}}}}
{{\resizebox{2.3in}{1.95in}{\includegraphics{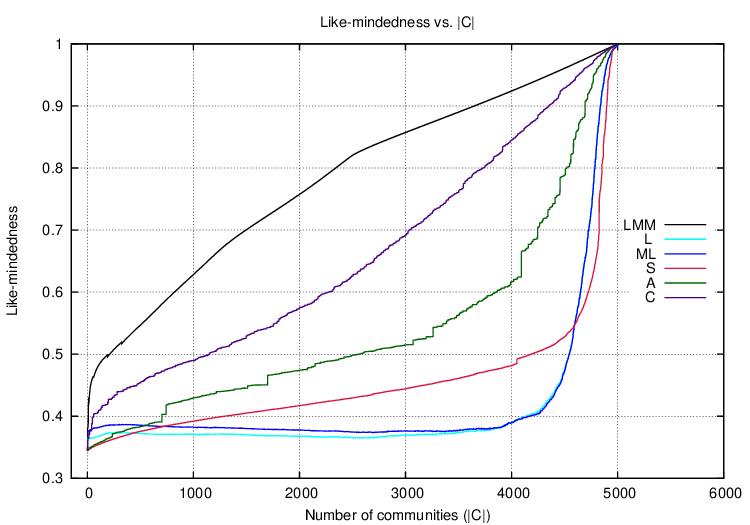}}}}
\caption{Comparison of modularity (top row) and like-mindedness (bottom row) scores on Filmtipset Filtered Networks $1$ (left) and $2$ (middle), and Twitter Filtered Network (right)}
\label{likemindedness}
}
\end{figure*}

We execute the Girvan-Newman algorithm, Louvain method, the single-linkage, average-linkage and complete linkage based hierarchical clustering algorithms, LMM and Modified Louvain methods on Filmtipset Filtered Networks $1$ and $2$. Among these, the Girvan-Newman algorithm and the Louvain method use only the edges of the networks. The remaining algorithms are executed using the rating and the (un)interested vectors separately. On Twitter Filtered Network, we execute all the algorithms mentioned above, except Girvan-Newman since the running time of this algorithm is too high on such a large graph. Note that the Louvain and the Modified Louvain methods terminate when a certain number of communities corresponding to the approximately highest modularity score is identified. All other algorithms used in this paper produce an hierarchy of communities, starting with $|V|$ communities and ending when there is only one community left, or vice-versa.

In Figure \ref{likemindedness}, the like-mindedness and modularity scores achieved by different algorithms are compared by plotting their values against $|C|$, the number of communities identified. In Table \ref{mod_values}, we tabulate the maximum modularity achieved by the algorithms on all $3$ datasets. In Table \ref{algo:time}, we summarize the running time of the algorithms considered in this paper. The notations we have used in these Figures and Tables are explained in Table \ref{table:symbols}. 

The key observation from these figures and tables is that our algorithm Like-mindedness Maximization outperforms all other algorithms (even Modified Louvain method) on \emph{Like-mindedness} metric. We also observe from Table \ref{mod_values} that it does not perform that well to obtain a community structure with high modularity. However, when the number of identified communities is large, we observe from Figure \ref{likemindedness} that our algorithm obtains a community structure with comparable (to other algorithms) modularity. The running time of our algorithm is 
$O({|V|}^2 \log |V|)$, which is much faster than the Girvan-Newman algorithm but slightly slower than other hierarchical clustering algorithms. We also note that the actual ratings (not just the data about whether a user has rated a movie or not) given by Filmtipset users does not give any significant advantage to the performance of the community detection algorithms. This is due to the fact that the similarity matrices of rating and (un)interested vectors of the user pairs are quite similar, e.g., having a cosine similarity of $0.9420$ and $0.9062$ for Filmtipset Filtered Network $1$ and $2$, respectively. Another interesting observation is that all the algorithms obtain higher like-mindedness and modularity scores in Filmtipset Filtered Network $2$ compared to Network $1$, due to it having higher homophily ratio. 

Let us conclude the paper with a few remarks on the potential applications of this work. Leskovek et al. \cite{viral} showed that community
finding is important in the context of viral marketing, in which a small group of users are targeted to promote a product in a community.
Since the community structure identified by our algorithm has high like-mindedness, it is likely that viral marketing would be more successful
among such like-minded groups of people. Also, the non-overlapping communities identified by our algorithm can be used as the core community structure in a network to create overlapping communities by adding additional nodes to each community. One drawback of our algorithm is its running time that is not suitable for very large networks. However, it can be used to identify the like-minded community structure among the most active and social users of
a network, the size of which may not be very large. Finally, it is an interesting open problem to design an algorithm that optimizes both the metrics modularity and like-mindedness of the identified community structure.

\section*{Acknowledgment}
This work has been supported by Ramanujan Fellowship, Department of Science and Technology, Government of India, grant number SR/S2/RJN-87/2011.

\end{document}